\documentclass{osa-article}

\journal{oe}

\usepackage{braket}
\usepackage{comment}

\begin{document}

\title{Generation of Schr\"{o}dinger cat states with Wigner negativity using continuous-wave low-loss waveguide optical parametric amplifier}

\author{Kan Takase,\authormark{1} Akito Kawasaki,\authormark{1} Byung Kyu Jeong,\authormark{1} Mamoru Endo,\authormark{1} Takahiro Kashiwazaki,\authormark{2} Takushi Kazama ,\authormark{2} Koji Enbutsu,\authormark{2} Kei Watanabe,\authormark{2} Takeshi Umeki,\authormark{2} Shigehito Miki,\authormark{3,4} Hirotaka Terai,\authormark{3} Masahiro Yabuno,\authormark{3} Fumihiro China,\authormark{3} Warit Asavanant,\authormark{1} Jun-ichi Yoshikawa,\authormark{1} and Akira Furusawa\authormark{1,5,*}}

\address{\authormark{1}Department of Applied Physics, School of Engineering, The University of Tokyo, 7-3-1 Hongo, Bunkyo-ku, Tokyo 113-8656, Japan\\
\authormark{2}NTT Device Technology Labs, NTT Corporation, 3-1, Morinosato Wakamiya, Atsugi, Kanagawa 243-0198, Japan\\
\authormark{3}Advanced ICT Research Institute, National Institute of Information and Communications Technology, 588-2 Iwaoka, Nishi-ku, Kobe, Hyogo 651-2492, Japan\\
\authormark{4}Graduate School of Engineering, Kobe University, 1-1 Rokkodai-cho, Nada-ku, Kobe, Hyogo 657-0013, Japan\\
\authormark{5}Optical Quantum Computing Research Team, RIKEN Center for Quantum Computing, 2-1 Hirosawa, Wako, Saitama 351-0198, Japan}

\email{\authormark{*}akiraf@ap.t.u-tokyo.ac.jp} 



\begin{abstract*}
Continuous-wave (CW) squeezed light is used in generation of various optical quantum states thus is a fundamental resource of fault-tolerant universal quantum computation using optical continuous variables. To realize a practical quantum computer, a waveguide optical parametric amplifier (OPA) is an attractive CW squeezed light source in terms of its THz-order bandwidth and suitability for modularization. The usages of a waveguide OPA in quantum applications thus far, however, are limited due to the difficulty of the generation of the squeezed light with a high purity. In this paper, we report the first observation of Wigner negativity of the states generated by a heralding method using a waveguide OPA. We generate Schr\"{o}dinger cat states at the wavelength of 1545 nm with Wigner negativity using a quasi-single-mode ZnO-doped periodically poled ${\rm LiNbO_3}$ waveguide module we developed. Wigner negativity is regarded as an important indicator of the usefulness of the quantum states as it is essential in the fault-tolerant universal quantum computation. Our result shows that our waveguide OPA can be used in wide range of quantum applications leading to a THz-clock optical quantum computer.
\end{abstract*}

\section{Introduction}
Quantum information processing (QIP) using optical continuous-variable (CV) is a one of the most promising candidate for realization of  fault-tolerant universal quantum computation \cite{Takeda2021}. In optical CV QIP, a continuous wave (CW) squeezed light source is a fundamental resource used for the generation of various quantum states. In general, because quantum states are very loss-sensitive, low-loss CW squeezed light sources are indispensable to quantum computation. In addition to losses, frequency bandwidth of the squeezed light sources also plays an important role in optical quantum computation. This is because the bandwidth of the squeezed light determines the upper limit of the clock frequency of the quantum computer, meaning that broadband squeezed light sources are desirable. CW squeezed light is widely generated from nonlinear processes such as spontaneous parametric down conversion (SPDC) using $\chi^{2}$ medium \cite{Wu1986,Polzik1992,Suzuki2021,Takeno2007,Vahlbruch2016} and four-wave mixing using atomic ensembles \cite{Slusher1985,McCormick2007,McCormick2008,Glorieux2011}. In optical QIP, SPDC based CW squeezed light source like an optical parametric oscillator (OPO) and an optical parametric amplifier (OPA) have long been used.

An OPO uses the SPDC process which is enhanced in an optical cavity. An OPO has been used in various experiments in CV QIP because it can generate CW squeezed light with a high purity \cite{Polzik1992,Suzuki2021,Takeno2007,Vahlbruch2016}. One prominent experiment using OPOs is the generation of large-scale quantum entangled states (two-dimensional cluster states) \cite{Warit2019,LarsenMikkel2019}, which enables scalable CV QIP by means of measurement-based quantum computation (MBQC) \cite{Raussendorf2001,Menicucci2006,Menicucci2011}. In order to perform fault-tolerant universal quantum computation using cluster states, it is necessary to prepare ancillary states with Wigner negativity \cite{Ohliger2010,Mari2012}. Many such states including photon number states \cite{Neergaard-Nielsen2007}, superposition of up-to three photons \cite{Yukawa2013}, and Schr\"{o}dinger cat states \cite{Neergaard-Nielsen2006,Wakui2007,Gerrits2010} have also been generated by using OPOs and photon detectors. Although an OPO has played an important role in the proof-of-principle experiments in quantum computation, it has a drawback that the generated squeezed light is narrowband due to the optical cavity.The reported bandwidth of an OPO is up to 2.5 GHz \cite{Ast2013} , despite the fact the phase-matching bandwidth of the $\chi^2$ medium itself is capable of generating THz-bandwidth squeezed light. For the development of a practical quantum computer, we need CW squeezed light source with a broader bandwidth because a narrow bandwidth limits the clock frequency of MBQC and the generation rate of ancillary states.

A waveguide OPA is promising as a broadband CW squeezed light source. Since a waveguide OPA uses SPDC without a cavity, it can generate THz-bandwidth squeezed light using the full range of the phase matching bandwidth of the $\chi^{2}$ medium. In a waveguide OPA, strong $\chi^{2}$ nonlinearity can be obtained in a single pass even for CW light because the pump light and the fundamental light interact over a long distance with high power density. Furthermore, a waveguide OPA is well suited for modularization and integration \cite{Montaut2017,Takanashi2020}, making it a realistic choice as a light source to be used in the practical quantum computers. Up until now, however, there are many difficulties in the generation of the high-purity squeezed light generated by the waveguide OPA: degradation from the anti-squeezed lights in the higher-order spatial modes, losses due to fabrication imperfections, and pump-light-induced loss. These factors limit the squeezing level in the conventional waveguide OPAs to about 2 dB \cite{Pysher2009,Mondain2019}. Hence, applications of a waveguide OPA to CV QIP remain limited compared to an OPO. To the best of our knowledge, generation of the quantum states with Wigner negativity using a waveguide OPA has not been achieved. Recently, we reported the realization of a dry-etched single-mode ZnO-doped periodically poled ${\rm LiNbO_3}$ (PPLN) waveguide with low loss and high pump durability \cite{Kashiwazaki2021}. A squeezing level of 6 dB was observed using this waveguide OPA module, indicating that it has much higher performance than conventional waveguides. Even more recently, we have realized a waveguide OPA module with even less loss by changing the fabrication method to mechanical sculpturing \cite{Kashiwazaki2022}. In this paper, we generate Schr\"{o}dinger cat states using this mechanically sculptured waveguide. The generated states are evaluated by homodyne measurements, and the Wigner negativity is observed without any loss correction. This result shows that our waveguide OPA module has sufficient performance to be applied to CV QIP, which is a significant step toward the realization of a THz-clocked quantum computer.

\section{Experimental details}

\subsection{\label{experiment1}Waveguide OPA module}

We use a quasi-single-mode ZnO-doped PPLN waveguide module we developed \cite{Kashiwazaki2022} as a CW squeezed light source. This module can generate high-purity CW squeezed light thanks to the three properties: high pump-durability, suppression of higher-order spatial modes, and low propagation loss.

In waveguide $\chi^{2}$ medium, strong pumping causes phenomena that degrade squeezed light such as scattering due to photorefractive effect \cite{Glass1978,Jackel2021} or photo-induced infrared light absorption \cite{Furukawa2021,Imlau2021}. Our waveguide is mechanically sculptured from a ZnO-doped PPLN wafar directly bonded to a ${\rm LiTaO_3}$ substrate, which has higher pump durability than waveguides fabricated by ion exchange or metal diffusion \cite{UmekiAug.}. In our waveguide, squeezed light is generated around the wavelength of 1545 nm in the first and second order spacial modes. The contamination by the antisqueezing of the second spatial mode is suppressed when the pump light is the first spatial mode with a power distribution close to Gaussian, because it hardly interacts with the second spatial mode at 1545 nm. We previously used dry-etching to fabricate the waveguides \cite{Kashiwazaki2021}. However, the dry-etched waveguide has 25$\%$ propagation loss for 45 mm length caused by the rough sidewalls of the waveguide due to the deposition of by-products generated during dry etching. Mechanically sculptured waveguides have smooth sidewalls and the propagation loss for 45-mm length is improved to 7$\%$. CW squeezed light generated by the mechanically sculptured waveguide was evaluated in an all-fiber system using optical amplification by another mechanically sculptured waveguide to improve the measurement signal-to-noise ratio \cite{Kashiwazaki2022}. As a result, a squeezing level of 6.3 dB was observed from the DC component to the 6 THz sideband while the squeezing level observed in a similar configuration with dry-etched waveguides was 3.2 dB \cite{Takanashi2020}. This result shows the potential of our mechanically sculptured waveguide for application in quantum technologies.

\subsection{\label{experiment2}Heralded generation of Schr\"{o}dinger cat states}

Schr\"{o}dinger cat states are superpositions of macroscopically distinguishable states, and in the particular case of the optical system, they are defined as a superpition of coherent states with opposite phases. A coherent state with amplitude $\alpha$ is given by $\ket{\alpha} = e^{(\alpha \hat{a}^{\dag}-\alpha^*\hat{a})}\ket{0},$ where $\hat{a}$ and $\hat{a}^{\dag}$ are annihilation and creation operators and $\ket{0}$ is a vacuum state. $\hat{a}$ and $\hat{a}^{\dag}$ satisfy $[\hat{a},\hat{a}^{\dag}]=1$. Cat states are given by
\begin{eqnarray}
\Ket{{\rm Cat}_{\phi}(\alpha)} &=& \frac{1}{N_{\phi, \alpha}} \left[ \ket{\alpha}+e^{i\phi}\ket{-\alpha} \right],
\end{eqnarray}
where $N_{\phi, \alpha}=\sqrt{2\left[ 1+\exp{(-2|\alpha|^2)\cos{\phi}} \right]}$. Cat state is a fundamental state in CV QIP because it can be regarded as a rough approximation of the Gottesman-Kitaev-Priskill (GKP) qubit \cite{Gottesman2001}, currently the most promising bosonic code for fault-tolerant universal CV MBQC, and is also the lowest-order case of the multi-component cat qubit \cite{Ralph2003,Mirrahimi2014,Albert2018}. The better approximations of GKP qubits and higher-order multi-component cat qubits can also be generated using cat states \cite{Vasconcelos2010,Weigand2018,Hastrup2020}.

Optical states with Wigner negativity, including cat states, are often generated by heralded scheme. The typical procedure of the heralded state generation is as follows. First, a quantum entangled state is generated by the squeezed light and linear optics. Then, photon detection is performed in the subsystem of the entangled state. This nonlinearity of the photon detection is essential as it is not possible to generate quantum states with Wigner negativity using only the squeezed light sources and linear optics. This nonlinearity is transmitted to the remaining unmeasured mode through the entanglement, and a state with Wigner negativity is generated.

In this paper, we perform the photon subtraction from squeezed light, which is a widely used  heralded scheme to generate the Schr\"{o}dinger cat states \cite{Dakna1997}. CW squeezed light is given by
\begin{eqnarray}
&&\hat{S}_r\Ket{\tilde{0}}=\exp{\left[ \frac{1}{2}\left( \hat{P}_r^{\dag}-\hat{P}_r \right) \right]}, \\
&&\hat{P}_r^{\dag} = \iint r(t_1-t_2) \hat{a}^{\dag}(t_1)\hat{a}^{\dag}(t_2)\ dt_1dt_2,
\end{eqnarray}
where $\Ket{\tilde{0}}$ is a multimode vacuum state, $\hat{a}(t), \hat{a}^{\dag}(t)$ are the instantaneous annihilation and creation operators satisfying $[\hat{a}(t), \hat{a}^{\dag}(t')]=\delta (t-t')$, and $r(t)$ is the correlation function of photon-pair generation satisfying $r(t)=r(-t)$. In the photon subtraction method, a squeezed light is slightly tapped and quantum entanglement is generated between the tapped channel (idler channel) and the other channel (signal channel). Optical filters are often inserted in the idler channel to purify the heralded states. When a photon is detected in the idler channel which has a filter with an impulse response $g(t)$, the heralded state in the signal channel is given by \cite{Yoshikawa2017}
\begin{eqnarray}
&&\hat{a}_{N(g^R)}\hat{S}_r\Ket{\tilde{0}} \propto \hat{S}_r \hat{a}_{N(g^R*r)}^{\dag}\Ket{\tilde{0}}, \\
&&\ \ \ \ \hat{a}_{h}^{\dag} = \int \ h(t)\hat{a}^{\dag}(t) \ dt,
\end{eqnarray}
where $N(\cdot)$ is a normalizing operation, $g^R(t)=g(-t)$, and $\hat{a}_{h}^{\dag}$ is a creation operator of mode $h$. The mode function $h(t)$ is normalized so that $\hat{a}_{h}^{\dag}$ satisfies $[\hat{a}_h,\hat{a}_{h}^{\dag}]=1$. When the CW squeezed light is much broader than the bandwidth of the optical filter, the heralded state is a single mode pure state whose mode function is given by
\begin{eqnarray}\label{eq:modefunction}
f(t) = N\left(g^R \right).
\end{eqnarray}
In this situation, the heralded state has only odd number photon components since the squeezed light has only even number photon components, one of the characteristic of the cat state with $\phi=\pi$. By appropriately choosing the experimental parameter, it is known that photon-subtracted squeezed light has a high fidelity to the cat state.

\subsection{\label{experiment3}Experimental apparatus and analysis}

\begin{figure}[t]
	\begin{center}
		\includegraphics[bb= 0 0 855 170,clip,width=0.95\textwidth]{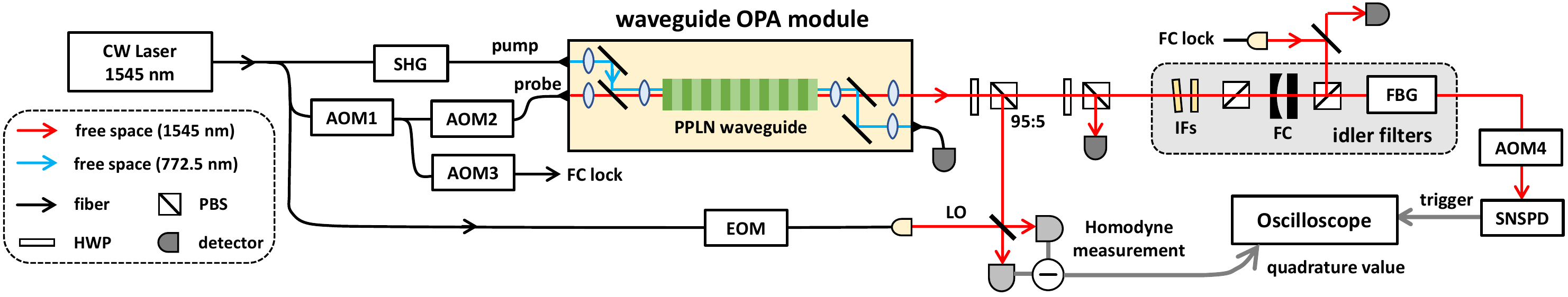}
	\end{center}
	\caption{Experimental setup. CW, Continuous Wave; AOM, Acousto-Optic-Modulator; SHG, Second Harmonic Generation; EOM, Electro-Optic-Modulator; OPA, Optical Parametric Oscillator; PBS, Polarization Beam Splitter; HWP, Half Wave Plate; LO, Local Oscillator; IF, Interference Filter; FC, Filter Cavity; FBG, Fiber Bragg Grating; SNSPD, Superconducting Nanostrip Photon Detector.}
\label{Fig:setup}
\end{figure}

Figure \ref{Fig:setup} shows the setup of the experiment. The main light source is a CW laser at 1545 nm. The waveguide OPA module is pumped with a 772.5 nm CW generated form a second-harmonic-generation module. The squeezed light is emitted into the free space, then it is separated into a signal channel and an idler channel by a half-wave plate and a polarization beamsplitter. The tapping ratio to the idler channel is set to 5$\%$. A homodyne detector is placed in the signal channel to evaluate the heralded states.

Since the bandwidth of the homodyne measurement device is limited to 200 MHz, the bandwidth of the states generated by photon subtraction must be within this range. In this experiment, we aim to generate a state with a bandwidth about 10 MHz by taking into account the clock of an existing quantum information processor \cite{Asavanant2021}. As we have shown in Eq. (\ref{eq:modefunction}), the mode property of the generated states in this experiment is determined only by the optical filters as the frequency range of the photon subtraction ($\sim$10 MHz) is much narrower than that of the squeezed light ($\sim$6 THz). The optical filter we constructed has a center wavelength of 1545 nm (=194 THz) and is consists of three types of filters: interference filter (HWHM=260 GHz), an fiber Bragg grating (HWHM=3.6 GHz), and a filter cavity (HWHM=8.2 MHz, FSR=8.5 GHz). Photon detections are performed by a superconducting nanostrip photon detector (SNSPD) \cite{Miki2017}. The SNSPD is installed into an adiabatic demagnetization refrigerator with an operational temperature of $\sim$500 mK. A detection efficiency and a dark count are 63$\%$ and 100 count per second (cps), respectively. The timing jitter of the SNSPD is about 100 ps, sufficient for the current experimental parameters.

The phase of the heralded states is controlled by a probe light. The probe light and lock light of the filter cavity need to be turned off while measurement of quantum states because they are so strong compared to quantum lights that the SNSPD easily saturates with them. We use acousto-optic modulators (AOMs) to switch between the control phase and the measurement phase with a period of 1.6 kHz. In the control phase, AOM1 to 3 in Fig. \ref{Fig:setup} is open but the AOM4 is closed to protect the SNSPD. In the measurement phase, AOM1 to 3 is closed and AOM4 is open to detect the photons tapped from the squeezed light.

The generated states are evaluated by measuring the quadrature given by 
\begin{eqnarray}
\hat{x}_{\theta} = \frac{\hat{a}e^{-i\theta}+\hat{a}^{\dag}e^{i\theta}}{\sqrt{2}} \ \ \ \ \ \ (\hbar=1).
\end{eqnarray}
In the following, we also use the notations $\hat{x}=\hat{x}_{\theta=0}$ and $\hat{p}=\hat{x}_{\theta=90}$. The value of $\theta$, corresponding to the phase of the local oscillator of homodyne measurement, can be switched by a waveguide phase modulator. We measure 20,000 events for each measurement bases: $\theta$ = 0, 30, 60, 90, 120, and 150 degrees, with the phase of the squeezed direction being 90 degree. After we collect the measurement results, we estimate the mode function of the heralded states by using the principal component analysis (PCA) on the time-correlation function of the electrical signals of the homodyne detector \cite{Abdi2021,MacRae2012,Morin2013}. The quadrature values are then obtained by integration of the electrical signals using the obtained mode. Finally, we perform quantum state tomography \cite{Lvovsky2009} using the quadrature values and estimate the Wigner functions of the heralded states.

\section{Results}
\begin{figure}[htbp]
	\begin{center}
		\includegraphics[bb= 0 0 430 280,clip,width=0.45\textwidth]{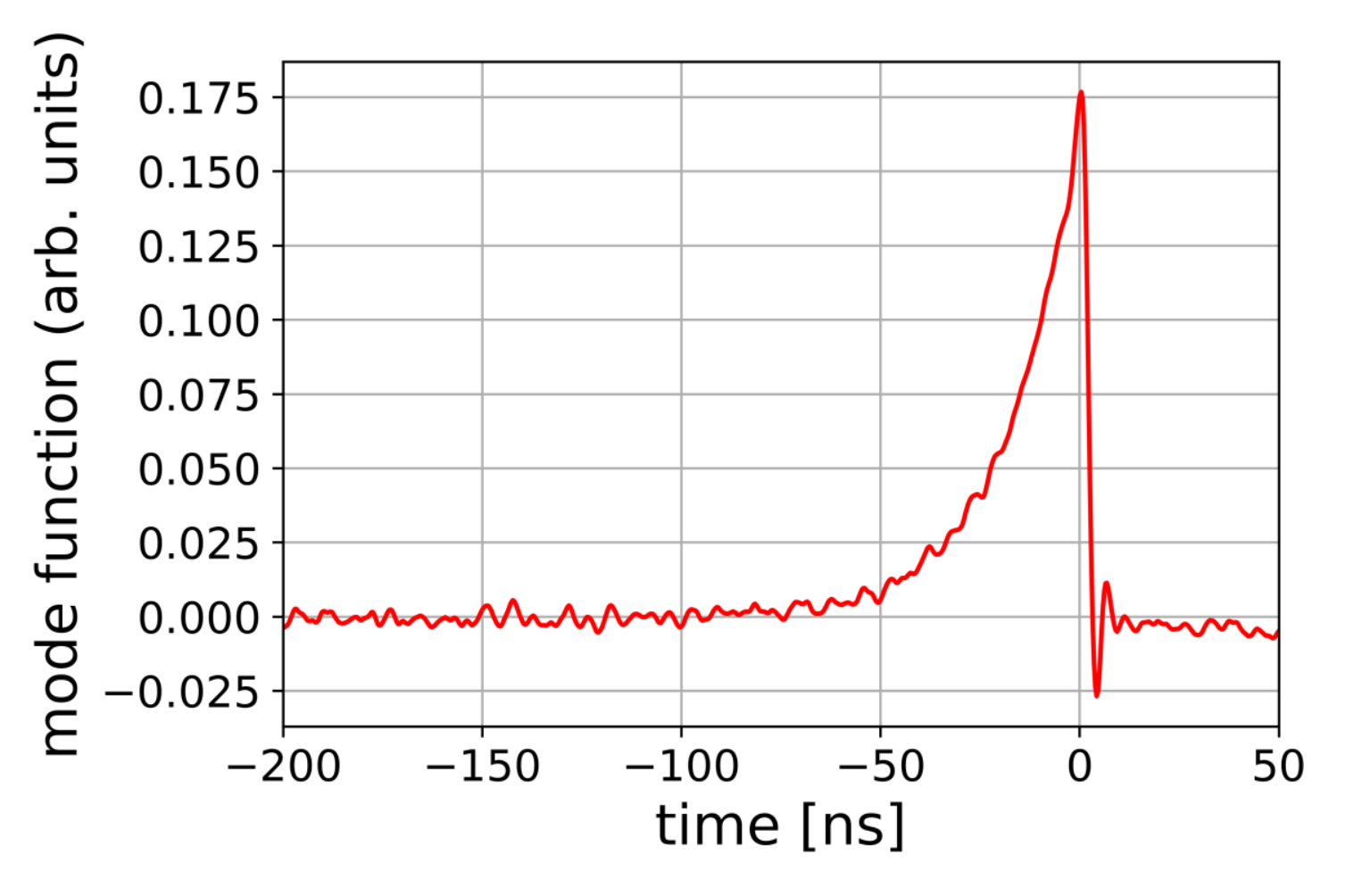}
	\end{center}
	\caption{Estimated modefunction of the heralded states. The exponentially rising shape is determined by the optical filter in the idler channel. The wide bandwidth of the waveguide OPA leads to the sharp function form compared to the previous researches using OPOs.}
\label{Fig:ps_modeunction}
\end{figure}

Experiments are conducted with four different pump power of 6 mW, 12 mW, 25 mW, and 50 mW. Figure \ref{Fig:ps_modeunction} shows the estimated mode function of the heralded states with the pump light of 6 mW. This mode function has an exponentially rising shape $e^{\gamma t}\Theta(-t)$, where $\Theta(t)$ is Heaviside step function and $\gamma = 2\pi \times 8.2$ MHz. This function is the time inverse of the normalized impulse response of the idler filter as expected in Eq. (\ref{eq:modefunction}). Similar mode functions are also observed for different pump light intensities. Although similar mode functions are also observed in the experiments using OPOs, the peak of the exponentially rising functions are usually rounded due to the bandwidth limitation of the OPOs \cite{Ogawa2016,Asavanant2017,Konno2021}. The sharp function form shown in Fig. \ref{Fig:ps_modeunction} indicates the broad bandwidth of the waveguide OPA.

\begin{figure}[htbp]
	\begin{center}
		\includegraphics[bb= 0 0 1170 580,clip,width=0.85\textwidth]{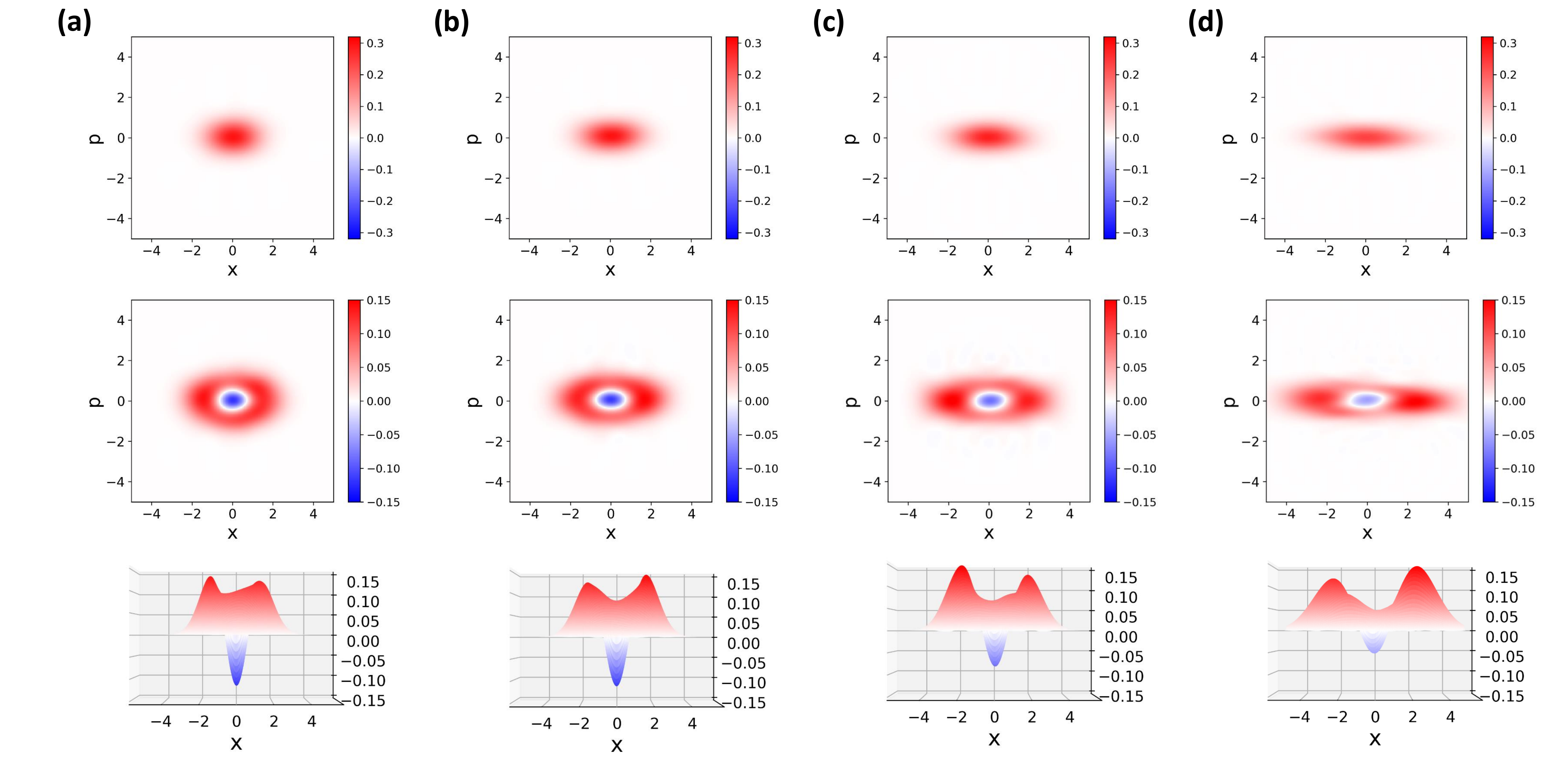}
	\end{center}
	\caption{The Wigner functions of the states without and with photon detection conditioning as for the mode estimated by PCA using the measurement results with photon detection. The top figure represents the state without conditioning. The pumping power is (a) 6 mW, (b) 12 mW, (c) 25 mW, and (d) 50mW.}
\label{Fig:Wigner}
\end{figure}

\begin{figure}[htbp]
  \begin{minipage}[b]{0.45\linewidth}
    \centering
    \captionsetup{width=.95\linewidth}
    \includegraphics[bb= 0 0 400 280, scale=0.4]{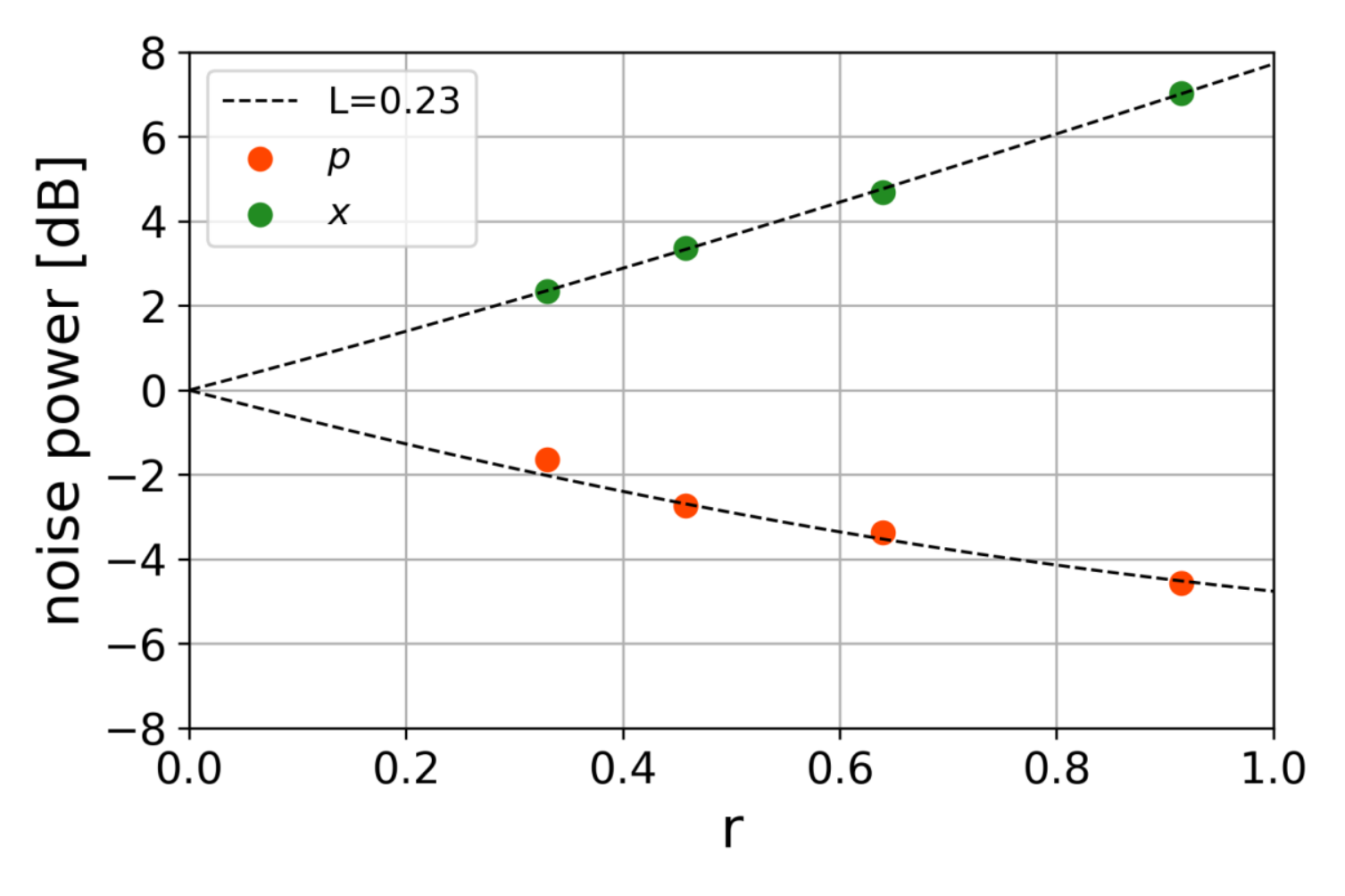}
    \caption{The quadrature variance of squeezed and antisqueezed direction about the states shown in the top of Fig. \ref{Fig:Wigner}. The experimental results are well explained by the theoretical line with 23$\%$ loss.}
    \label{Fig:squeezing}
  \end{minipage}
  \begin{minipage}[b]{0.45\linewidth}
    \centering
    \captionsetup{width=.95\linewidth}
    \includegraphics[bb= 0 0 400 280, scale=0.4]{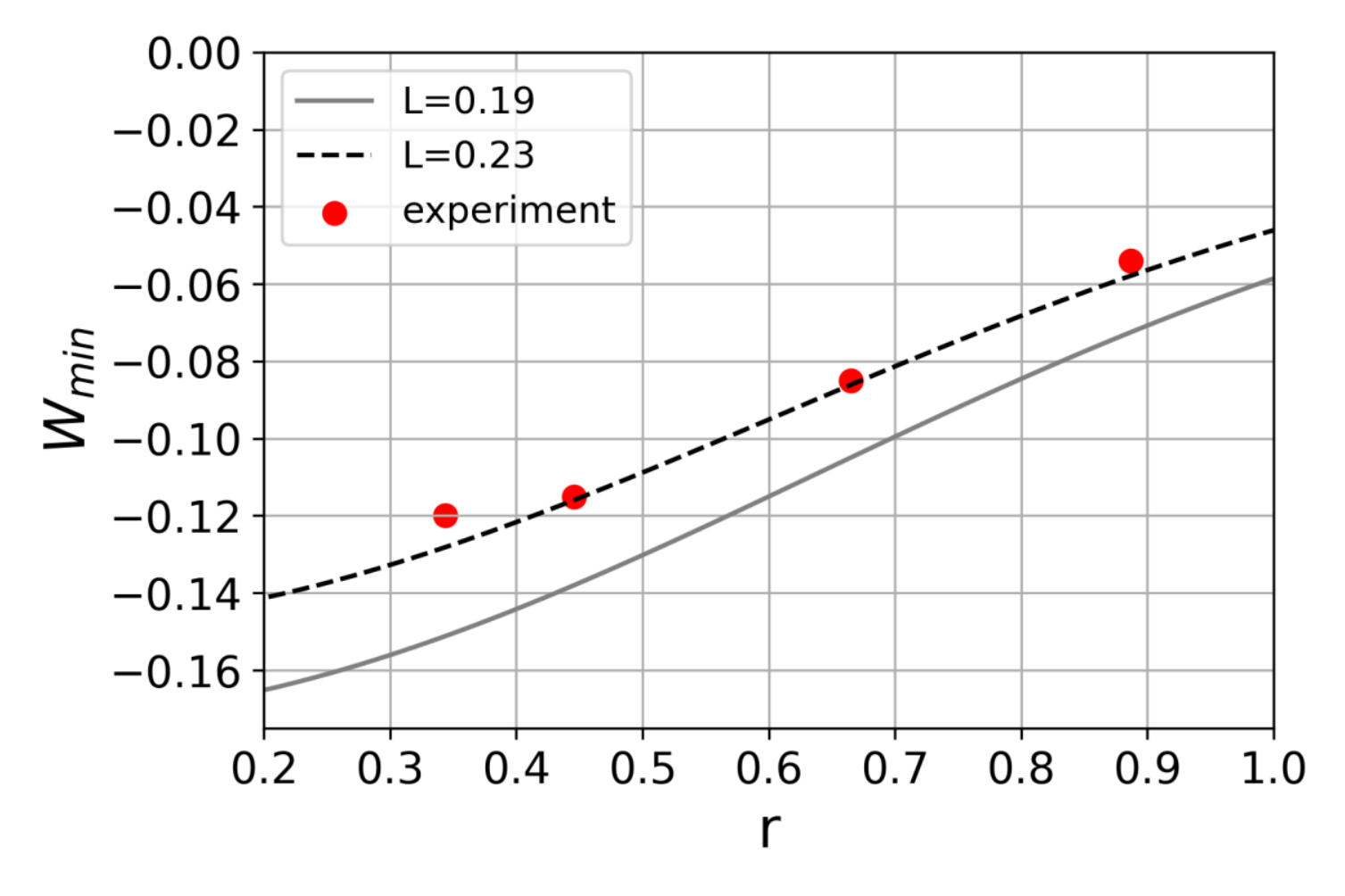}
    \caption{The minimum values of the Wigner functions about the state with photon detection conditioning. The experimental results are well explained by the theoretical line with 23$\%$ loss.}
    \label{Fig:wigner_negativity}
  \end{minipage}
\end{figure}

Figure \ref{Fig:Wigner} shows the Wigner functions of the states with and without conditioning by photon detection. We first consider Fig. \ref{Fig:Wigner} (a). The top figure is the Wigner function without photon detection, corresponding to the squeezed state emitted from the waveguide OPA. This Wigner function is a bivariate Gaussian function and has no negative values. We can see that the Wigner function is phase sensitive and the variance in the $p$-direction is small. The middle and bottom figures show the Wigner functions for the state conditioned by photon detection. It is worth noting these Wigner functions have negative values. This indicates that photon detection induces a highly non-classical state in the signal channel via quantum entanglement. The heralded state is phase-sensitive and has two peaks corresponding to the superposed coherent states in the $x$ direction, which indicates that the cat state has been successfully generated. For Figs. \ref{Fig:Wigner} (b)-(d), we can see that the squeezing becomes larger as the pump light becomes stronger. In addition, the negative value of the Wigner function degrades as the pump light increases. This is explained by two reasons. One reason is that the average photon number of the cat state increases as the squeezing level becomes high, making the state more sensitive to loss. Another reason is that the contamination by two photons subtraction increases with strong pump power.

The quadrature variance of squeezed light is given by squeezing parameter $r$ and loss $L$ as 
\begin{eqnarray}
\braket{\hat{x}_{\theta}^2}=\frac{1}{2}(1-L)[e^{2r}\cos^2\theta+e^{-2r}\sin^2\theta]+\frac{1}{2}L.
\end{eqnarray}
Figure \ref{Fig:squeezing} shows the value of $r$ and $L$ estimated from the squeezed states shown in Fig. \ref{Fig:Wigner}. The total loss is 23$\%$ and the loss budget is shown in Table \ref{tab:squeezed_loss}. Each item except for the OPA loss is directly measured. From the losses of these elements, the loss of the waveguide OPA module alone is estimated to be 11$\%$. The total loss of the homodyne measurement is 6$\%$, of which 2$\%$ is due to the mode mismatch between the squeezed light and the LO light emitted from the fiber collimator. This indicates that the waveguide OPA has a spatial mode suitable for connection to Gaussian modes and fiber modes.

Figure \ref{Fig:wigner_negativity} shows the minimum value of the Wigner functions of the generated cat states. The theoretical lines take into account the effect of two-photon subtraction and the loss of the heralded states. Experimental values are well explained by the theoretical line with 23$\%$ loss and the loss budget is shown in Table \ref{tab:cat_loss}. The main factor is a 19$\%$ loss of the signal channel, which excludes the effect of tapping for the photon subtraction from the loss shown in Table \ref{tab:squeezed_loss}. Secondly, fake counts of SNSPD have similar effect as losses on the negative value of Wigner functions which locate around the origin of the phase space. Fake counts are caused by both of squeezed light and  classical stray light. The former counts caused by detecting the photons of the squeezed light passing through the filter cavity sideband. About $0.1\%$ of these photons pass through the fiber Bragg grating and reach to the SNSPD. That leads to a loss of 3$\%$ independent of the pump power. The fake counts caused by classical light such as probe light and LO light are 210 cps in this experiment. This fake counts result in a loss of 0.1$\sim$1.7$\%$ because the whole counts are 12.6 kcps (with 6 mW pump), 33.6 kcps (12 mW), 75.6 cps (25 mW), and 174.3 kcps (50 mW). These count rates are much less than the saturation of the SNSPD. In addition, the electrical filtering of the homodyne signal adds a 2$\%$ loss. These losses explain the 23$\%$ loss. As described above, the properties of the generated state are consistent with the loss and other imperfections of the experimental system.

\begin{table}[htbp]
  \begin{minipage}[b]{0.46\linewidth}
    \centering
    \captionsetup{width=.95\linewidth}
    \begin{tabular}{l|l}
    \hline
    propagation loss & 3$\%$ \\
    tapping for photon subtraction&5$\%$ \\
    homodyne visibility&2$\%$ \\ 
    inefficiency of photodiodes\ \ \ \ \ &3$\%$\ \ \  \\ 
    circuit noise of the HD&1$\%$ \\
    waveguide OPA (estimated) &11$\%$ \\ 
    \hline
    total & 23$\%$ \\ \hline
    \end{tabular}
    \caption{\label{tab:squeezed_loss}Loss budget of the squeezed light. Total loss is estimated to be 23$\%$ as shown in \ref{Fig:squeezing}. The loss of each element except for the waveguide OPA is directly measured.}
  \end{minipage}
  \begin{minipage}[b]{0.46\linewidth}
    \centering
    \captionsetup{width=.95\linewidth}
    \begin{tabular}{l|l}
    \hline
    loss of the signal channel & 19$\%$ \\
    fake counts due to squeezed light\ &3$\%$ \\
    fake counts due to stray light&0.1$\sim$1.7$\%$ \\ 
    electrical signal processing &2$\%$\ \ \  \\ 
    \hline
    total & 23$\%$ \\ \hline
    \end{tabular}
    \caption{\label{tab:cat_loss}Loss budget of the generated cat states. The generated states are affected by the imperfection both of the signal channel and idler channel. Electrical signal processing also causes loss of the generated states.}
  \end{minipage}
\end{table}

\section{Conclusion}
In this paper, we have generated cat states with Wigner negativity using a waveguide OPA module we developed. In a previous research of cat state generation using a waveguide OPA, Wigner negativity have not been confirmed due to the performance of the light source \cite{Namekata2010}. The observation of Wigner negativity confirms the high performance of our waveguide OPA.

As a future outlook, development of a broadband homodyne measurement is essential for effective use of the broad bandwidth of a waveguide OPA. Broadband power detection of squeezed light is already achieved by using a waveguide OPA as an amplifier \cite{Takanashi2020,Kashiwazaki2022}. This technique can also be applied to a broadband homodyne measurement. With such a fast detection techniques, waveguide OPAs would lead to mainly two applications in CV QIP. One is the construction of a measurement-based information processor with a high clock frequency. Such a processor is much downsized than conventional processors as already discussed \cite{Kashiwazaki2021}, leading to long term stability required for a practical quantum computer. Another application is a high rate generation of ancillary states for CV MBQC. In the heralding scheme so far, generation rate of states requiring multi-photon detection is quite low, limiting the effective clock frequency of MBQC. The generation rate expected to be much improved by using a waveguide OPA as a light source because the bandwidth of the CW squeezed light corresponds to the repetition rate of the state generation. This paper shows that the waveguide OPA has reached a phase where it can be actively used in various applications in CV QIP, and is the first step toward the realization of a THz-clock optical quantum computer.

\begin{backmatter}
\bmsection{Funding}
Japan Society for the Promotion of Science KAKENHI (18H05207,20J10844,20K15187); Japan Science and Technology Agency (JPMJMS2064)

\bmsection{Acknowledgments}
The authors acknowledge supports form UTokyo Foundation and donations from Nichia Corporation of Japan. M.E. acknowledges supports from Research Foundation for Opto-Science and Technology.
A.K. acknowledges financial support from FoPM.

\end{backmatter}

\bibliography{OPA_cat}

\begin{thebibliography}{10}
\newcommand{\enquote}[1]{``#1''}

\bibitem{Takeda2021}
S.~Takeda and A.~Furusawa, \enquote{Toward large-scale fault-tolerant universal
  photonic quantum computing,} {\protect\JournalTitle{APL Photonics}}
  \textbf{4}, 060902 (2021).

\bibitem{Wu1986}
L.-A. Wu, H.~J. Kimble, J.~L. Hall, and H.~Wu, \enquote{Generation of squeezed
  states by parametric down conversion,} {\protect\JournalTitle{Phys. Rev.
  Lett.}} \textbf{57}, 2520--2523 (1986).

\bibitem{Polzik1992}
E.~S. Polzik, J.~Carri, and H.~J. Kimble, \enquote{Spectroscopy with squeezed
  light,} {\protect\JournalTitle{Phys. Rev. Lett.}} \textbf{68}, 3020--3023
  (1992).

\bibitem{Suzuki2021}
S.~Suzuki, H.~Yonezawa, F.~Kannari, M.~Sasaki, and A.~Furusawa, \enquote{7d{B}
  quadrature squeezing at 860nm with periodically poled {KTiOP${\rm O}_4$},}
  {\protect\JournalTitle{Appl. Phys. Lett.}} \textbf{89}, 061116 (2006).

\bibitem{Takeno2007}
Y.~Takeno, M.~Yukawa, H.~Yonezawa, and A.~Furusawa, \enquote{Observation of -9
  d{B} quadrature squeezing with improvement of phase stability in homodyne
  measurement,} {\protect\JournalTitle{Opt. Express}} \textbf{15}, 4321--4327
  (2007).

\bibitem{Vahlbruch2016}
H.~Vahlbruch, M.~Mehmet, K.~Danzmann, and R.~Schnabel, \enquote{Detection of 15
  d{B} squeezed states of light and their application for the absolute
  calibration of photoelectric quantum efficiency,}
  {\protect\JournalTitle{Phys. Rev. Lett.}} \textbf{117}, 110801 (2016).

\bibitem{Slusher1985}
R.~E. Slusher, L.~W. Hollberg, B.~Yurke, J.~C. Mertz, and J.~F. Valley,
  \enquote{Observation of squeezed states generated by four-wave mixing in an
  optical cavity,} {\protect\JournalTitle{Phys. Rev. Lett.}} \textbf{55},
  2409--2412 (1985).

\bibitem{McCormick2007}
C.~F. McCormick, V.~Boyer, E.~Arimondo, and P.~D. Lett, \enquote{Strong
  relative intensity squeezing by four-wave mixing in rubidium vapor,}
  {\protect\JournalTitle{Opt. Lett.}} \textbf{32}, 178--180 (2007).

\bibitem{McCormick2008}
C.~F. McCormick, A.~M. Marino, V.~Boyer, and P.~D. Lett, \enquote{Strong
  low-frequency quantum correlations from a four-wave-mixing amplifier,}
  {\protect\JournalTitle{Phys. Rev. A}} \textbf{78}, 043816 (2008).

\bibitem{Glorieux2011}
Q.~Glorieux, L.~Guidoni, S.~Guibal, J.-P. Likforman, and T.~Coudreau,
  \enquote{Quantum correlations by four-wave mixing in an atomic vapor in a
  nonamplifying regime: Quantum beam splitter for photons,}
  {\protect\JournalTitle{Phys. Rev. A}} \textbf{84}, 053826 (2011).

\bibitem{Warit2019}
W.~Asavanant, Y.~Shiozawa, S.~Yokoyama, B.~Charoensombutamon, H.~Emura, R.~N.
  Alexander, S.~Takeda, J.~Yoshikawa, N.~C. Menicucci, H.~Yonezawa, and
  A.~Furusawa, \enquote{Generation of time-domain-multiplexed two-dimensional
  cluster state,} {\protect\JournalTitle{Science}} \textbf{366}, 373--376
  (2019).

\bibitem{LarsenMikkel2019}
M.~V. Larsen, X.~Guo, C.~R. Breum, J.~S. Neergaard-Nielsen, and U.~L. Andersen,
  \enquote{Deterministic generation of a two-dimensional cluster state,}
  {\protect\JournalTitle{Science}} \textbf{366}, 369--372 (2019).

\bibitem{Raussendorf2001}
R.~Raussendorf and H.~J. Briegel, \enquote{A one-way quantum computer,}
  {\protect\JournalTitle{Phys. Rev. Lett.}} \textbf{86}, 5188--5191 (2001).

\bibitem{Menicucci2006}
N.~C. Menicucci, P.~van Loock, M.~Gu, C.~Weedbrook, T.~C. Ralph, and M.~A.
  Nielsen, \enquote{Universal quantum computation with continuous-variable
  cluster states,} {\protect\JournalTitle{Phys. Rev. Lett.}} \textbf{97},
  110501 (2006).

\bibitem{Menicucci2011}
N.~C. Menicucci, \enquote{Temporal-mode continuous-variable cluster states
  using linear optics,} {\protect\JournalTitle{Phys. Rev. A}} \textbf{83},
  062314 (2011).

\bibitem{Ohliger2010}
M.~Ohliger, K.~Kieling, and J.~Eisert, \enquote{Limitations of quantum
  computing with gaussian cluster states,} {\protect\JournalTitle{Phys. Rev.
  A}} \textbf{82}, 042336 (2010).

\bibitem{Mari2012}
A.~Mari and J.~Eisert, \enquote{Positive wigner functions render classical
  simulation of quantum computation efficient,} {\protect\JournalTitle{Phys.
  Rev. Lett.}} \textbf{109}, 230503 (2012).

\bibitem{Neergaard-Nielsen2007}
J.~S. Neergaard-Nielsen, B.~M. Nielsen, H.~Takahashi, A.~I. Vistnes, and E.~S.
  Polzik, \enquote{High purity bright single photon source,}
  {\protect\JournalTitle{Opt. Express}} \textbf{15}, 7940--7949 (2007).

\bibitem{Yukawa2013}
M.~Yukawa, K.~Miyata, T.~Mizuta, H.~Yonezawa, P.~Marek, R.~Filip, and
  A.~Furusawa, \enquote{Generating superposition of up-to three photons for
  continuous variable quantum information processing,}
  {\protect\JournalTitle{Opt. Express}} \textbf{21}, 5529--5535 (2013).

\bibitem{Neergaard-Nielsen2006}
J.~S. Neergaard-Nielsen, B.~M. Nielsen, C.~Hettich, K.~M^^c3^^b8lmer, and E.~S.
  Polzik, \enquote{Generation of a superposition of odd photon number states
  for quantum information networks,} {\protect\JournalTitle{Phys. Rev. Lett.}}
  \textbf{97}, 083604 (2006).

\bibitem{Wakui2007}
K.~Wakui, H.~Takahashi, A.~Furusawa, and M.~Sasaki, \enquote{Photon subtracted
  squeezed states generated with periodically poled {KTiOP${\rm O}_4$},}
  {\protect\JournalTitle{Opt. Express}} \textbf{15}, 3568--3574 (2007).

\bibitem{Gerrits2010}
T.~Gerrits, S.~Glancy, T.~S. Clement, B.~Calkins, A.~E. Lita, A.~J. Miller,
  A.~L. Migdall, S.~W. Nam, R.~P. Mirin, and E.~Knill, \enquote{Generation of
  optical coherent-state superpositions by number-resolved photon subtraction
  from the squeezed vacuum,} {\protect\JournalTitle{Phys. Rev. A}} \textbf{82},
  031802 (2010).

\bibitem{Ast2013}
S.~Ast, M.~Mehmet, and R.~Schnabel, \enquote{High-bandwidth squeezed light at
  1550 nm from a compact monolithic {PPKTP} cavity,}
  {\protect\JournalTitle{Opt. Express}} \textbf{21}, 13572--13579 (2013).

\bibitem{Montaut2017}
N.~Montaut, L.~Sansoni, E.~Meyer-Scott, R.~Ricken, V.~Quiring, H.~Herrmann, and
  C.~Silberhorn, \enquote{High-efficiency plug-and-play source of heralded
  single photons,} {\protect\JournalTitle{Phys. Rev. Applied}} \textbf{8},
  024021 (2017).

\bibitem{Takanashi2020}
N.~Takanashi, A.~Inoue, T.~Kashiwazaki, T.~Kazama, K.~Enbutsu, R.~Kasahara,
  T.~Umeki, and A.~Furusawa, \enquote{All-optical phase-sensitive detection for
  ultra-fast quantum computation,} {\protect\JournalTitle{Opt. Express}}
  \textbf{28}, 34916--34926 (2020).

\bibitem{Pysher2009}
M.~Pysher, R.~Bloomer, C.~M. Kaleva, T.~D. Roberts, P.~Battle, and O.~Pfister,
  \enquote{Broadband amplitude squeezing in a periodically poled {KTiOP${\rm
  O}_4$} waveguide,} {\protect\JournalTitle{Opt. Lett.}} \textbf{34}, 256--258
  (2009).

\bibitem{Mondain2019}
F.~Mondain, T.~Lunghi, A.~Zavatta, E.~Gouzien, F.~Doutre, M.~De~Micheli,
  S.~Tanzilli, and V.~D’Auria, \enquote{Chip-based squeezing at a telecom
  wavelength,} {\protect\JournalTitle{Photon. Res.}} \textbf{7}, A36--A39
  (2019).

\bibitem{Kashiwazaki2021}
T.~Kashiwazaki, N.~Takanashi, T.~Yamashima, T.~Kazama, K.~Enbutsu, R.~Kasahara,
  T.~Umeki, and A.~Furusawa, \enquote{Continuous-wave 6-d{B}-squeezed light
  with 2.5-{TH}z-bandwidth from single-mode {PPLN} waveguide,}
  {\protect\JournalTitle{APL Photonics}} \textbf{5}, 036104 (2021).

\bibitem{Kashiwazaki2022}
T.~Kashiwazaki, T.~Yamashima, N.~Takanashi, A.~Inoue, T.~Umeki, and
  A.~Furusawa, \enquote{Fabrication of low-loss quasi-single-mode {PPLN}
  waveguide and its application to a modularized broadband high-level
  squeezer,} {\protect\JournalTitle{Appl. Phys. Lett.}} \textbf{119}, 251104
  (2021).

\bibitem{Glass1978}
A.~M. Glass, \enquote{The photorefractive effect,}
  {\protect\JournalTitle{Optical Engineering}} \textbf{17}, 470--479 (1978).

\bibitem{Jackel2021}
J.~Jackel, A.~M. Glass, G.~E. Peterson, C.~E. Rice, D.~H. Olson, and J.~J.
  Veselka, \enquote{Damage‐resistant {L}i{N}b{O}{$_3$} waveguides,}
  {\protect\JournalTitle{Journal of Applied Physics}} \textbf{55}, 269--270
  (2021).

\bibitem{Furukawa2021}
Y.~Furukawa, K.~Kitamura, A.~Alexandrovski, R.~K. Route, M.~M. Fejer, and
  G.~Foulon, \enquote{Green-induced infrared absorption in {M}g{O} doped
  {L}i{N}b{O}{$_3$},} {\protect\JournalTitle{Appl. Phys. Lett.}} \textbf{78},
  1970--1972 (2021).

\bibitem{Imlau2021}
M.~Imlau, H.~Badorreck, and C.~Merschjann, \enquote{Optical nonlinearities of
  small polarons in lithium niobate,} {\protect\JournalTitle{Applied Physics
  Reviews}} \textbf{2}, 040606 (2021).

\bibitem{UmekiAug.}
T.~Umeki, O.~Tadanaga, and M.~Asobe, \enquote{Highly efficient wavelength
  converter using direct-bonded {PPZnLN} ridge waveguide,}
  {\protect\JournalTitle{IEEE Journal of Quantum Electronics}} \textbf{46},
  1206--1213 (Aug.).

\bibitem{Gottesman2001}
D.~Gottesman, A.~Kitaev, and J.~Preskill, \enquote{Encoding a qubit in an
  oscillator,} {\protect\JournalTitle{Phys. Rev. A}} \textbf{64}, 012310
  (2001).

\bibitem{Ralph2003}
T.~C. Ralph, A.~Gilchrist, G.~J. Milburn, W.~J. Munro, and S.~Glancy,
  \enquote{Quantum computation with optical coherent states,}
  {\protect\JournalTitle{Phys. Rev. A}} \textbf{68}, 042319 (2003).

\bibitem{Mirrahimi2014}
M.~Mirrahimi, Z.~Leghtas, V.~V. Albert, S.~Touzard, R.~J. Schoelkopf, L.~Jiang,
  and M.~H. Devoret, \enquote{Dynamically protected cat-qubits: a new paradigm
  for universal quantum computation,} {\protect\JournalTitle{New Journal of
  Physics}} \textbf{16}, 045014 (2014).

\bibitem{Albert2018}
V.~V. Albert, K.~Noh, K.~Duivenvoorden, D.~J. Young, R.~T. Brierley,
  P.~Reinhold, C.~Vuillot, L.~Li, C.~Shen, S.~M. Girvin, B.~M. Terhal, and
  L.~Jiang, \enquote{Performance and structure of single-mode bosonic codes,}
  {\protect\JournalTitle{Phys. Rev. A}} \textbf{97}, 032346 (2018).

\bibitem{Vasconcelos2010}
H.~M. Vasconcelos, L.~Sanz, and S.~Glancy, \enquote{All-optical generation of
  states for “{E}ncoding a qubit in an oscillator”,}
  {\protect\JournalTitle{Opt. Lett.}} \textbf{35}, 3261--3263 (2010).

\bibitem{Weigand2018}
D.~J. Weigand and B.~M. Terhal, \enquote{Generating grid states from
  {S}chr\"odinger-cat states without postselection,}
  {\protect\JournalTitle{Phys. Rev. A}} \textbf{97}, 022341 (2018).

\bibitem{Hastrup2020}
J.~Hastrup, J.~S. Neergaard-Nielsen, and U.~L. Andersen, \enquote{Deterministic
  generation of a four-component optical cat state,}
  {\protect\JournalTitle{Opt. Lett.}} \textbf{45}, 640--643 (2020).

\bibitem{Dakna1997}
M.~Dakna, T.~Anhut, T.~Opatrn\'y, L.~Kn\"oll, and D.-G. Welsch,
  \enquote{Generating {S}chr\"odinger-cat-like states by means of conditional
  measurements on a beam splitter,} {\protect\JournalTitle{Phys. Rev. A}}
  \textbf{55}, 3184--3194 (1997).

\bibitem{Yoshikawa2017}
J.~Yoshikawa, W.~Asavanant, and A.~Furusawa, \enquote{Purification of photon
  subtraction from continuous squeezed light by filtering,}
  {\protect\JournalTitle{Phys. Rev. A}} \textbf{96}, 052304 (2017).

\bibitem{Asavanant2021}
W.~Asavanant, B.~Charoensombutamon, S.~Yokoyama, T.~Ebihara, T.~Nakamura, R.~N.
  Alexander, M.~Endo, J.~Yoshikawa, N.~C. Menicucci, H.~Yonezawa, and
  A.~Furusawa, \enquote{Time-domain-multiplexed measurement-based quantum
  operations with 25-{MH}z clock frequency,} {\protect\JournalTitle{Phys. Rev.
  Appl.}} \textbf{16}, 034005 (2021).

\bibitem{Miki2017}
S.~Miki, M.~Yabuno, T.~Yamashita, and H.~Terai, \enquote{Stable,
  high-performance operation of a fiber-coupled superconducting nanowire
  avalanche photon detector,} {\protect\JournalTitle{Opt. Express}}
  \textbf{25}, 6796--6804 (2017).

\bibitem{Abdi2021}
H.~Abdi and L.~J. Williams, \enquote{Principal component analysis,}
  {\protect\JournalTitle{WIREs Comp Stat}} \textbf{2}, 433--459 (2021).

\bibitem{MacRae2012}
A.~MacRae, T.~Brannan, R.~Achal, and A.~I. Lvovsky, \enquote{Tomography of a
  high-purity narrowband photon from a transient atomic collective excitation,}
  {\protect\JournalTitle{Phys. Rev. Lett.}} \textbf{109}, 033601 (2012).

\bibitem{Morin2013}
O.~Morin, C.~Fabre, and J.~Laurat, \enquote{Experimentally accessing the
  optimal temporal mode of traveling quantum light states,}
  {\protect\JournalTitle{Phys. Rev. Lett.}} \textbf{111}, 213602 (2013).

\bibitem{Lvovsky2009}
A.~I. Lvovsky and M.~G. Raymer, \enquote{Continuous-variable optical
  quantum-state tomography,} {\protect\JournalTitle{Rev. Mod. Phys.}}
  \textbf{81}, 299--332 (2009).

\bibitem{Ogawa2016}
H.~Ogawa, H.~Ohdan, K.~Miyata, M.~Taguchi, K.~Makino, H.~Yonezawa,
  J.~Yoshikawa, and A.~Furusawa, \enquote{Real-time quadrature measurement of a
  single-photon wave packet with continuous temporal-mode matching,}
  {\protect\JournalTitle{Phys. Rev. Lett.}} \textbf{116}, 233602 (2016).

\bibitem{Asavanant2017}
W.~Asavanant, K.~Nakashima, Y.~Shiozawa, J.-I. Yoshikawa, and A.~Furusawa,
  \enquote{Generation of highly pure {S}chr\"{o}dinger's cat states and
  real-time quadrature measurements via optical filtering,}
  {\protect\JournalTitle{Opt. Express}} \textbf{25}, 32227--32242 (2017).

\bibitem{Konno2021}
S.~Konno, A.~Sakaguchi, W.~Asavanant, H.~Ogawa, M.~Kobayashi, P.~Marek,
  R.~Filip, J.~Yoshikawa, and A.~Furusawa, \enquote{Nonlinear squeezing for
  measurement-based non-gaussian operations in time domain,}
  {\protect\JournalTitle{Phys. Rev. Appl.}} \textbf{15}, 024024 (2021).

\bibitem{Namekata2010}
N.~Namekata, Y.~Takahashi, G.~Fujii, D.~Fukuda, S.~Kurimura, and S.~Inoue,
  \enquote{Non-gaussian operation based on photon subtraction using a
  photon-number-resolving detector at a telecommunications wavelength,}
  {\protect\JournalTitle{Nature Photonics}} \textbf{4}, 655--660 (2010).

\end{thebibliography}






\end{document}